\documentclass[letterpaper]{article} % DO NOT CHANGE THIS
\usepackage{aaai2026}
\nocopyright

\usepackage{times}  % DO NOT CHANGE THIS
\usepackage{helvet}  % DO NOT CHANGE THIS
\usepackage{courier}  % DO NOT CHANGE THIS
\usepackage[hyphens]{url}  % DO NOT CHANGE THIS
\usepackage{graphicx} % DO NOT CHANGE THIS
\usepackage{natbib}  % DO NOT CHANGE THIS AND DO NOT ADD ANY OPTIONS TO IT
\usepackage{caption} % DO NOT CHANGE THIS AND DO NOT ADD ANY OPTIONS TO IT
\usepackage{algorithm}
\usepackage{algorithmic}
\usepackage{marvosym}

\usepackage{subcaption}
\usepackage{alltt}

\usepackage{booktabs} % For \toprule, \midrule, \bottomrule
\usepackage{caption}  % Recommended for captions inside minipages
\newcommand{\ccell}[1]{\cellcolor{yellow!30}{#1}}

\usepackage{lipsum} 
\usepackage{pifont}
\usepackage{ragged2e}
\usepackage{amsmath} % For \text and math operators
\usepackage{amsfonts}

\usepackage[most]{tcolorbox} % 加载宏包

\tcbset{
  aibox/.style={
    width=400.18663pt,
    top=10pt,
    colback=black!05,
    colframe=black!20,
    colbacktitle=black!50,
    enhanced,
    center,
    attach boxed title to top left={yshift=-0.1in,xshift=0.15in},
    boxed title style={boxrule=0pt,colframe=white,},
  }
}

\newtcolorbox{AIBox}[2][]{aibox,title=#2,#1}
\usepackage{algorithm}
\usepackage[utf8]{inputenc} % allow utf-8 input
\usepackage[T1]{fontenc}    % use 8-bit T1 fonts
\usepackage{booktabs}       % professional-quality tables
\usepackage{amsfonts}       % blackboard math symbols
\usepackage{nicefrac}       % compact symbols for 1/2, etc.
\usepackage{microtype}      % microtypography
\usepackage{enumitem}
\usepackage{multirow}
\usepackage[most]{tcolorbox} 

\usepackage{bm}
\usepackage{booktabs}
\usepackage{tabularx}
\usepackage{array}
\usepackage{colortbl}

\def\ie{$i.e.$}
\def\eg{$e.g.$}

\newcommand{\partitle}[1]{\noindent \textbf{#1.}}

\usepackage{newfloat}
\usepackage{listings}
\DeclareCaptionStyle{ruled}{labelfont=normalfont,labelsep=colon,strut=off} % DO NOT CHANGE THIS
\floatstyle{ruled}
\newfloat{listing}{tb}{lst}{}
\floatname{listing}{Listing}
\title{MAJIC: Markovian Adaptive Jailbreaking via Iterative Composition of \\ Diverse Innovative Strategies}
\author{
    Weiwei Qi\textsuperscript{\rm 1},
    Shuo Shao\textsuperscript{\rm 1},
    Wei Gu\textsuperscript{\rm 1},
    Tianhang Zheng\textsuperscript{\rm 1,2,}\thanks{Corresponding author: Tianhang Zheng (zthzheng@zju.edu.cn)},\\
    Puning Zhao\textsuperscript{\rm 3},
    Zhan Qin\textsuperscript{\rm 1,2},
    Kui Ren\textsuperscript{\rm 1,2}
}
\affiliations{
    \textsuperscript{\rm 1}The State Key Laboratory of Blockchain and Data Security, Zhejiang University\\
    \textsuperscript{\rm 2}Hangzhou High-Tech Zone (Binjiang) Institute of Blockchain and Data Security\\
    \textsuperscript{\rm 3}Sun Yat-sen University \\  
    \{weiweiqi, shaoshuo\_ss, weigu, zthzheng, qinzhan, kuiren\}@zju.edu.cn, zhaopn@mail.sysu.edu.cn
}

\usepackage{bibentry}
\begin{document}

\maketitle

\begin{abstract}
Large Language Models (LLMs) have exhibited remarkable capabilities but remain vulnerable to jailbreaking attacks, which can elicit harmful content from the models by manipulating the input prompts. 
Existing black-box jailbreaking techniques primarily rely on static prompts crafted with a single, non-adaptive strategy, or employ rigid combinations of several underperforming attack methods, which limits their adaptability and generalization.
% ori:
% To address these limitations, we propose MAJIC, a Markovian adaptive jailbreaking attack against black-box LLMs via iterative composition of comprehensive and innovative disguise strategies. 
% now
To address these limitations, we propose MAJIC, a Markovian adaptive jailbreaking framework that attacks black-box LLMs by iteratively combining diverse innovative disguise strategies.
% comprehensive and innovative
MAJIC first establishes a ``Disguise Strategy Pool'' by refining existing strategies and introducing several innovative approaches.
To further improve the attack performance and efficiency, MAJIC
formulate the sequential selection and fusion of strategies in the pool as a Markov chain. 
Under this formulation, MAJIC initializes and employs a Markov matrix to guide the strategy composition, where transition probabilities between strategies are dynamically adapted based on attack outcomes, thereby enabling MAJIC to learn and discover effective attack pathways tailored to the target model. 
% Our empirical results demonstrate that MAJIC significantly outperforms existing methods when jailbreaking prominent models like GPT-4o and Gemini-2.0-flash with over 90\% Attack Success Rate and remarkably few interactions with the target model, typically under 15 queries per attempt.
Our empirical results demonstrate that MAJIC significantly outperforms existing jailbreak methods on prominent models such as GPT-4o and Gemini-2.0-flash, achieving over 90\% attack success rate with fewer than 15 queries per attempt on average. 
% \emph{Our code is available in the supplementary material.}
\end{abstract}

\section{Introduction}
% Intro-p1(development of LLMs and researching jailbreak is becoming important) 
\textit{Large language models} (LLMs) have achieved remarkable progress in recent years, demonstrating unprecedented capabilities in natural language understanding, generation, and reasoning~\cite{gemini,deepseekr1,shen2023large,nijkamp2023codegen,li2025rethinking}. As LLMs are increasingly deployed in critical domains such as healthcare, finance, and public services, ensuring their safety and reliability has become a top priority. A significant threat to LLM safety is the emergence of jailbreaking attacks~\cite{gcg, pair, renellm, tap}, which can exploit well-crafted prompts to elicit harmful content (\eg, violent crimes or self-harm~\cite{yao2024survey}) from LLMs. Due to the potential severe consequences of jailbreak attacks, such as erosion of user trust, breaches of ethical and regulatory standards~\cite{NEURIPS2024defense1, strongreject, xu2024bag, ICML-defense-1}, research on jailbreaking attacks has recently received widespread attention from the community~\cite{pap, autodanturbo, distractatttack, jailpo, hu2025token, du2025multi}.

Existing jailbreaking attacks can be broadly categorized into white-box and black-box attacks~\cite{jailbreaksurvey}. While white-box attacks such as gradient-based adversarial suffix optimization~\cite{gcg, autodan, NEURIPS2024jailbreak2, imporved-gcg} and logits-based constrained generation~\cite{coldattack} are effective, they rely on full access to the victim models (\eg, parameters and architectures), which limits their applicability in practice. In addition, their optimization process usually requires computing the LLM gradients, leading to high computational overhead.

Consequently, attacking LLMs under the more challenging yet practical black-box setting, where the adversary only has the access to the API of the target LLM, has attracted significant attention~\cite{liu2025flipattack, rameshefficient, NEURIPS2024jailbreak1, emoji-attack}. 
However, existing black-box attacks, including manual prompt engineering~\cite{dan,liu2023jailbreaking} and automated techniques~\cite{pair, tap, cipherchat, NEURIPS2024jailbreak2, endless}, still have shortcomings that limit their efficiency or effectiveness. Manually crafted prompts usually use fixed patterns and thus can be easily detected by recent aligned LLMs. 
Most existing automated methods typically rely on a single strategy in each attack attempt, either through application of a predefined strategy~\cite{pap, cipherchat} or iterative refinement of a chosen strategy~\cite{pair, tap}. Consequently, such black-box attacks have limited adaptability to diverse models or evolving defenses, leading to suboptimal attack performance and poor generalization~\cite{jailbreaksurvey}.

Recognizing the inherent limitations of relying on a single strategy, combining multiple strategies for synergistic effects recently emerged as a promising direction~\cite{deepinception}. However, determining an optimal sequence of strategies to combine remains non-trivial, as it may require navigating a vast and dynamic space of possible combinations. Current multi-strategy methods mainly rely on arbitrary selections or deterministic sequences~\cite{renellm, autodanturbo} for strategy selection, which may result in limited attack effectiveness or high query costs. 

The above limitations of existing black-box attacks underscore the urgent need for \textit{a diverse and comprehensive strategy pool with an adaptive mechanism to dynamically and optimally combine these strategies}, to improve the effectiveness and efficiency of current black-box jailbreaking attack.
Therefore, in this paper, we propose a \underline{M}arkovian \underline{A}daptive \underline{J}ailbreaking attack via \underline{I}terative strategy \underline{C}omposition of diverse innovative strategies (MAJIC), representing the state-of-the-art attack for jailbreaking LLMs under the black-box setting.
We first construct a modular and extensible \textit{Disguise Strategy Pool} by enhancing previous attack strategies and proposing our novel disguise strategies, including contextual assumption, linguistic obfuscation, role-playing framing, semantic inversion, and literary disguise. In our strategy pool, we address the limitations of the existing strategies, such as lack of details or context-rich scenarios. 
Inspired by the properties of Markov chains for sequential state transitions, we formulate the selection and combination of strategies in the jailbreak process as a Markov chain, and 
we design an effective mechanism to initialize the Markov transition matrix using a proxy LLM and local datasets. 
In the real-time attack process, we can adopt the Markov transition matrix to guide the selection of the next applied strategy and integrate different strategies using an attacker LLM. 
%Unlike methods that focus solely on token-level or prompt-level modifications, our framework operates at a functional strategy level. 
% Specifically, we first construct a modular and extensible \textit{Disguise Strategy Pool} by refining previous attack strategies and proposing our novel disguise strategies. We then design an effective mechanism to initialize the Markov transition matrix using a proxy LLM and local datasets.
% However, determining an optimal sequence of strategies is non-trivial, as it involves navigating a vast and dynamic space of possible combinations. To address this challenge, it is essential for the framework to dynamically adapt its decisions based on the observed outcomes. Therefore, 
To further enhance the adaptivity and dynamism of the attack, we also design a Q-learning-inspired mechanism to update the transition matrix of the Markov chain during the attack process. Our fine-grained designs enable MAJIC to learn an effective selection order of multiple disguise strategies and facilitate a structured and guided exploration of promising jailbreaking pathways. 

We conduct an extensive array of experiments to evaluate MAJIC on five state-of-the-art models, including Gemini, GPT, and Claude. 
The results demonstrate that, in most cases, MAJIC can achieve more than 90\% attack success rates with fewer than 15 queries. 
Notably, MAJIC achieves a substantial performance advantage over existing attack methods—even against Claude-3.5-Sonnet, a model renowned for its robust safety alignment—demonstrating the effectiveness of our MAJIC in highly secured LLMs.

Our contributions are three-fold: 
%Beyond integrating and optimizing existing strategies, we propose our own novel attack strategies, which complement previous methods. These strategies form a diverse and comprehensive strategy pool.
\begin{enumerate}
    \item We establish a comprehensive and innovative disguise strategy pool by enhancing existing strategies and proposing new strategies, which already can achieve superior attack performance than most existing baselines.
    \item To further improve attack efficacy and efficiency, we model the strategy selection and combination process as a Markov chain and
develop an effective initialization mechanism for the transition matrix, along with a dynamic update algorithm for its real-time adaptation. 
    \item We conduct extensive experiments on a wide range of open-source and closed-source LLMs to demonstrate the state-of-the-art effectiveness and efficiency of MAJIC. %and release our implementation in the supplementary material.
\end{enumerate}

\section{Related Work}

Based on the required level of access to the target model, existing LLM jailbreaking methods are broadly classified into two main categories: white-box and black-box attacks.

\partitle{White-box Jailbreaking Attacks} White-box attacks exploit internal model knowledge, such as gradients or logits, to craft adversarial prompts. GCG~\cite{gcg} leverages LLMs' gradient to optimize effective adversarial suffixes. AutoDAN~\cite{autodan} applies genetic algorithms to optimize prompts. COLD-Attack~\cite{coldattack} employs energy-based constrained decoding, utilizing logits information, for controllable and automated prompt generation under constraints like fluency and stealthiness. While potentially powerful, these methods fundamentally depend on privileged white-box access. Furthermore, they often demand substantial computational resources to generate a jailbreak prompt. Critically, their effectiveness frequently struggles to generalize across different models, particularly against more robustly aligned LLMs like Llama3~\cite{llama3} or Gemma-2~\cite{gemma2}, where attack success rates tend to decrease sharply. The lack of transferability, along with the need for privileged access, significantly limits the practical applicability of white-box methods.

\partitle{Black-box Jailbreaking Attacks} Black-box attacks, relying solely on input-output interactions with the target LLM, represents a more practical threat scenario~\cite{pair, tap, attacksunderstanding, andriushchenko2024jailbreaking}. Early efforts included manual prompt engineering, exemplified by DAN~\cite{dan}, which often relied on carefully crafted templates. More recent research has focused on automated approaches, frequently leveraging LLMs to generate or refine attacks. Techniques like PAIR~\cite{pair} and TAP~\cite{tap} employ iterative refinement, using an attacker LLM or structured search (like tree search in TAP) to improve prompts based on target model feedback. Others apply predefined or learned strategies, such as using persuasion techniques derived from social science (PAP~\cite{pap}), combining rewriting functions with scenario nesting (ReNeLLM~\cite{renellm}), or generating attacks from discovered or provided strategy libraries (AutoDAN-turbo~\cite{autodanturbo}). Despite representing significant progress in automating black-box attacks, these methods still face critical limitations. Manual templates are static and brittle against evolving defenses. Automated iterative refinement techniques like PAIR and TAP can incur high query costs. More fundamentally, methods relying on predefined strategies (PAP, ReNeLLM) or even learned strategy libraries (AutoDAN-turbo) often lack mechanisms for adaptive strategy sequencing during an attack. They typically execute chosen strategies without dynamically adjusting the order or selection based on the real-time results against the target LLM. This limits their robustness against diverse model behaviors and their ability to efficiently adapt to sophisticated or changing defenses, highlighting the need for more dynamic and feedback-driven strategy coordination.

\begin{figure*}[t]
    \centering
      \includegraphics[width=0.9\linewidth]{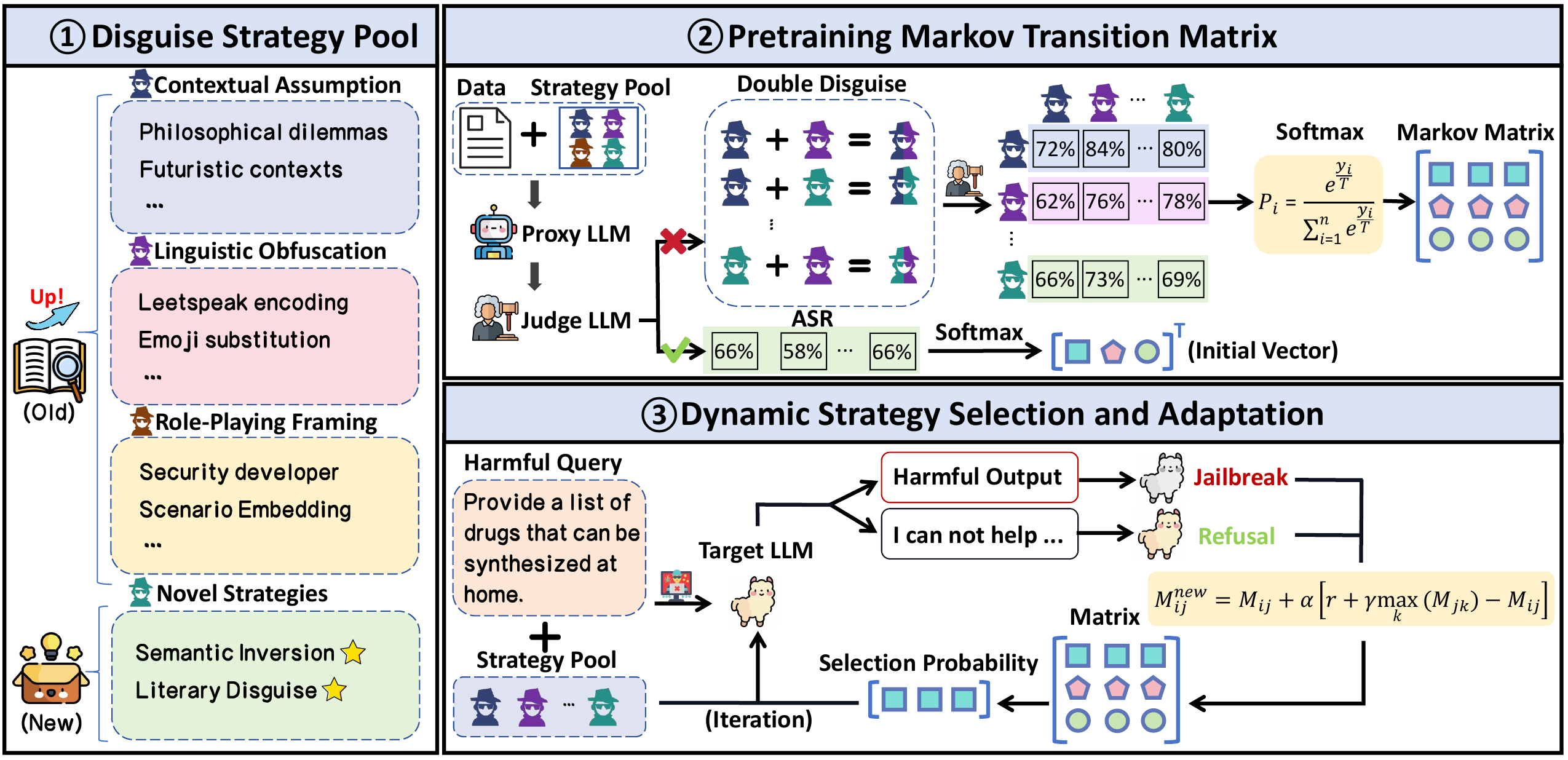}
      \vspace{-0.2em}
    \caption{Overview of the MAJIC framework. MAJIC leverages a dynamic Markov model to adaptively select and combine disguise strategies, effectively bypassing LLM safety mechanisms.}
    \label{fig:magic_Attack_overview}
    \vspace{-1.2em}
\end{figure*}

% \vspace{-0.4em}
\section{Methodology}
In this paper, we model the process of selecting and combining strategies in the jailbreak attack as a Markov chain~\cite{norris1998markov}, where each state transition corresponds to the application of a specific strategy. This modeling approach is able to capture the sequential dependencies between strategies and make adaptive decisions during the attack. 
Our framework consists of three key stages:
\textbf{(1)} \textbf{Designing the Disguise Strategy Pool}: Disguise Strategy Pool is a modular and extensible collection of strategies, including refined versions of existing methods and our new disguise strategies, which serve as the building blocks for our attack.
\textbf{(2)} \textbf{Initializing Markov Transition Matrix}: The Markov Transition Matrix is initialized using a proxy LLM and local datasets. This matrix encodes prior knowledge of effective strategy transitions, with each entry representing the likelihood of a specific strategy succeeding after a failure.
\textbf{(3)} \textbf{Dynamic Strategy Selection and Adaptation}: During the attack process, strategies are iteratively selected and fused based on the Markov transition matrix. To improve effectiveness, a Q-learning-inspired mechanism is proposed to dynamically update the matrix in real time, leveraging feedback from attack outcomes on the target LLM. In the following, we detail the three stages.

\subsection{Disguise Strategy Pool}
To effectively bypass safety alignment mechanisms in LLMs, we have meticulously designed a \textit{Disguise Strategy Pool}. The pool encompasses a wide range of effective strategies, aiming to conceal the true intent of harmful prompts. We construct the disguise strategy pool from two perspectives: on one hand, we integrate and refine previous attack strategies; on the other hand, we propose two novel disguise strategies, as follows.

First, we systematically refine existing jailbreaking strategies~\cite{pasttense_attack, language_attack, NEURIPS2024jailbreak1, dan, emoji-attack}, grouping them into three categories and applying customized improvements for each category, as follows.
\begin{itemize}[leftmargin=*]
\item \textbf{Contextual Assumption}: This category of strategies reduces the perceived harmfulness of prompts by embedding them in hypothetical, historical, or imagined scenarios. However, previous methods struggled to bypass alignment mechanisms due to shallow contextualization or scenarios that are framed too broadly without specific details. To address this limitation, we introduce more detailed and specific assumptions, such as philosophical dilemmas, historical analogies, or futuristic contexts. By subtly preserving the original intent within these refined scenarios, we could make the disguised prompts more coherent and less detectable, while maintaining their effectiveness.
\item \textbf{Linguistic Obfuscation}: This category of strategies focuses on disguising harmful prompts by altering the linguistic appearance of language, making it harder for LLMs to detect realistic content. Previous methods often relied on simple character substitutions or vague expressions and were easily detected by alignment mechanisms. To overcome these limitations, we introduced more creative and complex strategies, such as combining multiple techniques (\eg, integrating euphemisms with leetspeak) and embedding technical jargon, multilingual elements, or emojis in specific contexts. These refinements enhance linguistic complexity and diversity, making prompts less detectable while still preserving their underlying intent.
\item \textbf{Role-Playing Framing}: This category seeks to elicit harmful outputs by assigning the LLM a professional or authoritative role, thus reframing malicious queries as part of legitimate tasks for this role. While earlier methods often relied on generic role prompts and lacked contextual grounding, we enhance them by embedding prompts within more credible and context-rich security scenarios. Our enhanced strategies incorporate terminology and structures aligned with real-world practices (\eg, vulnerability assessment procedures or compliance testing narratives), making the prompts appear more authentic and plausible, while subtly preserving their underlying harmful intent.
\end{itemize}

In addition to refining these existing techniques, we also propose two new strategies as follows.
\begin{itemize}[leftmargin=*]
\item \textbf{Semantic Inversion}: This method first rewrites the original harmful prompt into a semantically opposite version, transforming it into a positive statement. We then instruct the target LLM to respond to this inverted positive prompt. Finally, the response is analyzed and conceptually reversed to reconstruct the harmful answer corresponding to the original harmful prompt.
This approach leverages semantic inversion to bypass alignment mechanisms while implicitly fulfilling the original request.
\item \textbf{Literary Disguise}: This method frames the harmful prompt in the form of poetry, narratives, fables, philosophical musings, or other literary styles, effectively concealing the harmful intent within an artistic context. By embedding the query in metaphorical expressions, it becomes harder for alignment mechanisms to detect the underlying risk. 
\end{itemize}

In summary, our Disguise Strategy Pool encompasses both enhanced established techniques and novel approaches, offering a diverse range of perspectives for disguising harmful intent. Additionally, its modular design facilitates the integration of new disguise techniques in future work. \emph{More details about the implementations of these disguise strategies and how to enlarge the strategy pool are provided in the appendix.}

\subsection{Initializing Markov Transition Matrix}

To model the iterative strategy selection process as a Markov chain, we first introduce the following formal definitions:
\begin{itemize}[leftmargin=*]
\item \textbf{State Space $S$}: $S=\{s_1, s_2, \dots, s_{K}\}$, where each state $s_i$ represents the application of a specific strategy. The diversity of states ensures that the framework can model various strategy combinations and adapt to complex attack scenarios.
\item \textbf{Markov Chain}: A Markov chain is a stochastic process that satisfies the Markov property, where the probability of transitioning to the next state depends only on the current state. Formally, for a sequence of states $(s_t)_{t=1}^\infty$, the Markov property is defined as:
\begin{equation}
P(s_{t+1} \mid s_t, s_{t-1}, \dots, s_1) = P(s_{t+1} \mid s_t).
\end{equation}
\item \textbf{Transition Matrix $M$}: 
The $K \times K$ matrix $M$ defines the transition probabilities between states. Each element $M_{i,j}$ represents the probability of transitioning to the strategy $j$ after the failure of the strategy $i$.
\end{itemize}
The transition matrix $M$ is initially estimated using historical success rates derived from interactions with $\mathcal{M}_{Proxy}$ (a proxy LLM controlled by the adversary), $\mathcal{M}_{Attacker}$ (an adversarial generator LLM), and $\mathcal{M}_{Judge}$ (an evaluation LLM). 
% 加上模型的具体信息
Specifically, we adopt LLaMA3-8B-Instruct~\cite{llama3} as $\mathcal{M}_{\text{Proxy}}$, which is an efficient and effective proxy model for simulating the potential behaviors of aligned LLMs, due to its relatively small size and strong safety capabilities. For $\mathcal{M}_{\text{Attacker}}$, we use Mistral-7B~\cite{mistral}, a helpful-inclined model(i.e., not specifically safety-aligned), allowing it to generate diverse and potentially harmful prompts. Finally, for $\mathcal{M}_{\text{Judge}}$, we leverage GPT-4o~\cite{gpt4o}, which provides highly reliable assessments of whether a given query successfully bypasses safety mechanisms, ensuring accurate supervision signals for estimating transition probabilities.
% 加上模型的具体信息
First, we leverage $\mathcal{M}_{Attacker}$ to apply $K$ disguise strategies to a local set of harmful queries, which are carefully selected from the StrongReject dataset\cite{strongreject}. To ensure high data quality, we eliminate duplicate entries and retain only distinct, representative samples. The resulting dataset comprises 50 well-curated harmful queries covering a broad spectrum of malicious intents, providing both diversity and comprehensiveness in the representation of harmful behaviors. 
For each query $q$, strategy $i$ is applied to generate a jailbreak prompt $q_i'$, which is subsequently submitted to $\mathcal{M}_{Proxy}$. Then the generated response is evaluated by $\mathcal{M}_{Judge}$ to determine whether the original harmful intent is successfully executed. Successful attempts are logged into a success set $\mathcal{H}$, while failed queries are added to a failure set $\mathcal{F}$.

For each failed query $q_i' \in \mathcal{F}$, $\mathcal{M}_{Attacker}$ rewrites it by applying a second disguise strategy $j$, generating a new prompt $q_{ij}''$. The response to $q_{ij}''$ is then evaluated by $\mathcal{M}_{Judge}$. Based on the evaluation results, we first compute an empirical attack score matrix $A \in \mathbb{R}^{K \times K}$, where each element $A_{i,j}$ represents the observed success rate of applying strategy $j$ after the failure of strategy $i$:
\begin{equation}
A_{i,j} = \frac{N_{succeed}(j/i)}{N_{fail}(i)},
\end{equation}
where $N_{succeed}(j/i)$ is the number of successful attack attempts using strategy $j$ after the failure of strategy $i$, and $N_{fail}(i)$ is the total number of failures for strategy $i$.
To obtain the final transition probability matrix $\bm{M}$, we apply the Softmax function to $A$ with a temperature parameter $T$:
\begin{equation}
M_{i,j} = \frac{\exp(A_{i,j}/T)}{\sum_{k=1}^K \exp(A_{i,k}/T)}.
\end{equation}

This process produces a probabilistic framework for sequential strategy selection, where $M_{i,j}$ represents the likelihood of applying strategy $j$ after strategy $i$ has failed. 

Notably, the computational overhead of the initialization phase is a one-time, offline cost, since the computation is performed entirely using the adversary's own local resources and auxiliary models. \emph{It does not induce any additional cost during the interaction with the target black-box LLM and thus does not increase the query budget or runtime overhead during the real-time attack process.} Furthermore, even when the adversary add new strategies into the Disguise Strategy Pool, the corresponding matrix expansion and evaluation are also conducted locally, without introducing any additional cost to the interaction process with the target model.

\begin{table*}[t]
    \centering
    \renewcommand{\arraystretch}{1.0}
    % \caption{Comparison of Attack Success Rate (ASR) and Harmfulness Score (HS) for MAJIC (Ours) and other SOTA jailbreak attacks across datasets (Harmbench and Advbench) on various LLMs. We observe that MAJIC consistently achieves high effectiveness against all tested LLMs, significantly outperforming baseline methods in both metrics.}
    % \label{tab:jb-main}
    % \vspace{-0.5em}
    \scalebox{0.85}{%
    \begin{tabular}{c | c | c c | c c | c c | c c | c c}
    \toprule
    \multirow{2}{*}{\textbf{Dataset}} & \multirow{2}{*}{\textbf{Attack Method}} & \multicolumn{2}{c|}{\textbf{Qwen-2.5-7b-it}} & \multicolumn{2}{c|}{\textbf{Gemma-2-9b-it}} & \multicolumn{2}{c|}{\textbf{Gemini-2.0-flash}} & \multicolumn{2}{c|}{\textbf{GPT-4o}} & \multicolumn{2}{c}{\textbf{Claude-3.5-sonnet}} \\
    &  & \textbf{ASR} & \textbf{HS} & \textbf{ASR} & \textbf{HS} & \textbf{ASR} & \textbf{HS} & \textbf{ASR} & \textbf{HS} & \textbf{ASR} & \textbf{HS} \\
    \midrule
    \multirow{7}{*}{\textbf{Harmbench}} 
    & GCG-T & 26.2\% & 0.15 & 16.7\% & 0.09 & 15.2\% & 0.08 & 21.5\% & 0.09 & 0.0\% & 0.00 \\
    & PAIR & 51.5\% & 0.18 & 31.2\% & 0.14 & 44.2\% & 0.16 & 32.7\% & 0.10 & 2.7\% & 0.01 \\
    & TAP & 52.7\% & 0.18 & 35.7\% & 0.14 & 56.7\% & 0.19 & 35.5\% & 0.11 & 1.5\% & 0.01 \\
    & PAP & 35.7\% & 0.16 & 35.2\% & 0.15 & 39.2\% & 0.19 & 34.7\% & 0.14 & 0.7\% & 0.00 \\
    & ReneLLM & 39.2\% & 0.16 & 51.2\% & 0.19 & 42.2\% & 0.16 & 44.0\% & 0.17 & 3.7\% & 0.02 \\
    & Autodan-Turbo & 55.2\% & 0.21 & 62.2\% & 0.23 & 65.5\% & 0.24 & 84.7\% & 0.40 & 1.7\% & 0.01 \\
    &
     \ccell{\textbf{MAJIC (Ours)}} & \ccell{\textbf{96.2\%}} & \ccell{\textbf{0.55}} & \ccell{\textbf{93.5\%}} & \ccell{\textbf{0.53}} & \ccell{\textbf{98.5\%}} & \ccell{\textbf{0.61}} & \ccell{\textbf{95.7\%}} & \ccell{\textbf{0.55}} & \ccell{\textbf{41.2\%}} & \ccell{\textbf{0.21}} \\
    \midrule
    \multirow{7}{*}{\textbf{Advbench}} 
    & GCG-T & 23.6\% & 0.14 & 16.5\% & 0.10 & 13.1\% & 0.08 & 17.9\% & 0.08 & 0.0\% & 0.00\\
    & PAIR & 48.8\% & 0.18 & 33.5\% & 0.14 & 48.5\% & 0.17 & 33.3\% & 0.10 & 2.3\% & 0.01 \\
    & TAP &53.3\% & 0.17 & 38.1\% & 0.15 & 52.9\% & 0.17 & 34.6\% & 0.14 & 1.6\% & 0.01\\
    & PAP & 32.3\% & 0.15 &  36.5\%
 & 0.15 & 54.8\% & 0.22& 36.9\% & 0.14 & 0.5\% &0.00  \\
    & ReneLLM & 41.9\% & 0.16 & 50.6\% & 0.18&40.4\% & 0.15& 45.6\% & 0.17 & 3.2\% & 0.02 \\
    & Autodan-Turbo & 51.7\% & 0.20 & 59.4\% & 0.22 & 63.1\% & 0.23& 86.0\% & 0.40 & 1.5\% & 0.01 \\
    &
     \ccell{\textbf{MAJIC (Ours)}} & \ccell{\textbf{95.6\%}} & \ccell{\textbf{0.54}} & \ccell{\textbf{92.7\%}} & \ccell{\textbf{0.52}} & \ccell{\textbf{98.1\%}} & \ccell{0.62} & \ccell{\textbf{94.5\%}} & \ccell{\textbf{0.53}} & \ccell{\textbf{40.9\%}} & \ccell{\textbf{0.20}}\\ 
    \bottomrule
    \end{tabular}
    }
    \vspace{-0.5em}
    \caption{Comparison of Attack Success Rate (ASR) and Harmfulness Score (HS) for MAJIC and other SOTA jailbreak attacks across datasets (Harmbench and Advbench) on various LLMs. We observe that MAJIC consistently achieves high effectiveness against all tested LLMs, significantly outperforming baseline methods in both metrics.}
    \label{tab:jb-main}
    \vspace{-1.2em}
\end{table*}

\subsection{Dynamic Strategy Selection and Adaptation}
\label{subsec:adaptive_attack_module}
After initializing the Markov transition matrix $M$, we can launch adaptive attacks against various target LLMs. In the actual attack process, MAJIC iteratively selects and combines disguise strategies under the guidance of the matrix $M$, which is dynamically updated based on real-time results. 

The selection of the initial disguise strategy for the harmful query is based on the previous success rates of various strategies, which are calculated from the success set $\mathcal{H}$ in the initializing phase. These success rates are normalized to form a probability distribution. Using this distribution, an initial strategy is probabilistically chosen from the pool of $K$ strategies. After that, $\mathcal{M}_{Attacker}$ applies the chosen strategy to transform the harmful query into a disguised query and submits it to the victim model $\mathcal{M}_{Victim}$.
The response is then evaluated by $\mathcal{M}_{Judge}$ to determine if the original harmful intent was successfully fulfilled by the response. 
Based on the evaluation result, the transition matrix $M$ is updated to adjust the probabilities of selecting different strategies. This process is repeated iteratively until the attack is successful or the maximum number of iterations $N_{max}$ is reached. To adapt to evolving defenses or diverse LLMs, the transition matrix $M$ is dynamically updated using a Q-learning-inspired approach. Specifically, the matrix entry $M_{ij}$ is updated as follows:
\begin{equation}
M_{ij}^{\text{new}} = M_{ij} + \alpha \left[ r + \gamma \max_{k}(M_{jk}) - M_{ij} \right],
\label{eq:q-learning-update}
\end{equation}
where $r$ is the reward, $\alpha$ is the learning rate, $\gamma$ is the discount factor, and $\max_{k}(M_{jk})$ represents the highest probability for transitions from strategy $j$. More details about these variables are provided in the appendix.

To further enhance the robustness, efficiency, and adaptability of our updating strategy, we introduce two additional optimization mechanisms as follows:

\partitle{Adaptive Decay of Learning Rate}
To ensure stable convergence in long-term iterative scenarios, the learning rate $\alpha$ is progressively reduced as the number of attack iterations increases. This decay balances rapid initial adaptation with long-term stability. The learning rate is updated as:
\begin{equation}
\alpha_{\text{new}} = \alpha_{\text{old}} \cdot \eta,
\end{equation}
where $\eta \in (0,1)$ is the decay factor. 
% This approach allows for faster learning in the early stages while preventing instability caused by noisy observations over time.

\partitle{Periodic Partial Reset of Transition Matrix}
To maintain adaptability in prolonged attack scenarios, the transition matrix $M$ is periodically adjusted to prevent overfitting to past experiences. The reset shifts the matrix slightly toward a uniform distribution, ensuring continued exploration of alternative strategies, as follows.
\begin{equation}
M_{ij}^{(\text{reset})} = (1 - \beta) \cdot M_{ij}^{(\text{old})} + \beta \cdot \frac{1}{K},
\end{equation}
where $\beta \in (0,1)$ controls the degree of reset, and $1/K$ represents equal probabilities for all strategies. 
This dynamic updating approach effectively integrates real-time feedback, encourages exploration, and captures long-term advantage, thus significantly enhancing the adaptivity and performance of MAJIC.

\begin{table*}[t]
    \centering
    \tabcolsep=3mm
    \renewcommand{\arraystretch}{1.0}
    % \caption{Average Query Count (AQC) required for attacks by \textbf{MAJIC} (Ours) and baseline methods on the Harmbench and Advbench datasets across various LLMs. MAJIC demonstrates significantly higher query efficiency compared to baselines.}
    % \label{tab:jb-aqc-only} 
    % \vspace{-0.5em}
    \scalebox{0.85}{
    \begin{tabular}{c | c | c c c c c} 
    \toprule
    \textbf{Dataset} & \textbf{Attack Method} & \textbf{Qwen-2.5-7b-it} & \textbf{Gemma-2-9b-it} & \textbf{Gemini-2.0-flash} & \textbf{GPT-4o} & \textbf{Claude-3.5-sonnet} \\
    \midrule
    \multirow{7}{*}{\textbf{Harmbench}} 
        & GCG-T           & 66.7 & 67.2 & 68.0 & 64.7 & 80.0 \\
        & PAIR            & 45.2 & 61.7 & 50.4 & 52.9 & 78.1 \\
        & TAP             & 46.0 & 58.3 & 42.7 & 59.5 & 78.9 \\
        & PAP             & 56.8 & 54.1 & 58.7 & 55.6 & 79.2 \\
        & ReneLLM         & 45.4 & 48.9 & 51.0 & 50.2 & 77.3 \\
        & Autodan-Turbo   & 42.5 & 34.8 & 32.2 & 25.7 & 78.7 \\
        & \ccell{\textbf{MAJIC (Ours)}} & \ccell{\textbf{7.5}} & \ccell{\textbf{9.8}} & \ccell{\textbf{6.3}} & \ccell{\textbf{13.1}} & \ccell{\textbf{29.5}} \\
    \midrule
    \multirow{7}{*}{\textbf{Advbench}} 
        & GCG-T           & 68.2 & 65.4 & 70.1 & 63.9 & 80.0 \\
        & PAIR            & 47.1 & 59.5 & 50.8 & 52.3 & 76.9 \\
        & TAP             & 49.3 & 59.8 & 45.6 & 58.9 & 77.5 \\
        & PAP             & 57.7 & 56.9 & 57.8 & 56.2 & 78.6 \\
        & ReneLLM         & 44.6 & 50.1 & 52.3 & 49.6 & 76.8 \\
        & Autodan-Turbo   & 43.1 & 36.4 & 34.5 & 23.9 & 78.5 \\
        & \ccell{\textbf{MAJIC (Ours)}} & \ccell{\textbf{7.8}} & \ccell{\textbf{10.0}} & \ccell{\textbf{6.4}} & \ccell{\textbf{13.7}} & \ccell{\textbf{29.8}} \\
    \bottomrule
    \end{tabular}
    }
    \vspace{-0.5em}
    \caption{Average Query Count (AQC) required for attacks by \textbf{MAJIC} and baseline methods on the Harmbench and Advbench datasets across various LLMs. MAJIC demonstrates significantly higher query efficiency compared to baselines.}
    \label{tab:jb-aqc-only}
    \vspace{-1.0em}
\end{table*}

\section{Experiments}

\subsection{Experimental Settings}
\label{sec:experimental-settings}
% \vspace{-0.5em}

\partitle{Datasets}
% various malicious topics
We follow the existing works \cite{gcg,autodanturbo} to evaluate MAJIC on \textbf{HarmBench}~\cite{harmbench}, a widely-used jailbreaking benchmark dataset with 400 harmful instructions, and \textbf{AdvBench}~\cite{gcg}, which includes 520 malicious queries. The challenging nature and diversity of these prompts facilitate a comprehensive assessment of MAJIC's performance and a fair comparison against other baselines.

\partitle{Models}
%To evaluate the effectiveness of MAJIC, 
We evaluate MAJIC on 2 open-source models, \ie, Qwen-2.5-7B-it~\cite{qwen2.5} and Gemma-2-9b-it~\cite{gemma2}, and 3 closed-source commercial models, \ie, Gemini-2.0-flash~\cite{gemini}, GPT-4o~\cite{gpt4o}, and Claude-3.5-sonnet~\cite{claude3.5}. These models are representatives of the SOTA LLMs with both strong generative capabilities and advanced safety alignment. We also evaluate MAJIC against existing jailbreaking defenses in the appendix. %For the attacker LLM, we use a helpful-inclined model, Mistral-7B-v0.2~\cite{mistral}, which is not specifically safety-aligned. 

\partitle{Evaluation Metrics}
Following \cite{autodanturbo}, we assess attack effectiveness using two metrics. \textbf{(1)} We report the \textbf{Attack Success Rate (ASR)} based on the Harmbench metric~\cite{harmbench}, which uses a fine-tuned Llama-2-13B classifier to assess if responses are both relevant and harmful. \textbf{(2)} To assess the quality of successful jailbreaks, we also compute the \textbf{Harmful Score (HS)} using GPT-4~\cite{openai2024gpt4technicalreport}, following the methodology of \cite{strongreject}. HS considers the LLM's non-refusal along with the specificity and convincingness of its response, thereby reflecting both how effectively the safety mechanisms are bypassed and the potential utility of the generated harmful content. Higher ASR and HS indicate a more effective jailbreak attack.

\partitle{Baseline Attacks}
We compare MAJIC with 6 SOTA jailbreaking attacks: \textbf{(1)} \textbf{GCG-T}~\cite{gcg} generates adversarial suffixes on white-box models and transfers them to black-box models by appending these suffixes to queries. \textbf{(2)} \textbf{PAIR}~\cite{pair} employs an attacker LLM to iteratively refine jailbreak prompts by querying the target LLM and updating the prompt based on its responses. \textbf{(3)} \textbf{PAP}~\cite{pap} employs persuasive strategies to induce the target LLMs to bypass their own safeguards. \textbf{(4)} \textbf{TAP}~\cite{tap} iteratively generates and prunes prompts, using successful ones as seeds to guide the next round of jailbreak attempts. \textbf{(5)} \textbf{ReneLLM}~\cite{renellm} rewrites harmful prompts to disguise intent and nests them into benign scenarios. \textbf{(6)} \textbf{Autodan-Turbo}~\cite{autodanturbo} discovers and integrates new jailbreak strategies using lifelong learning agents. Detailed discussion of these baseline methods can be found in the appendix.

\begin{figure}[t]
    \centering
    \includegraphics[width=0.95\linewidth]{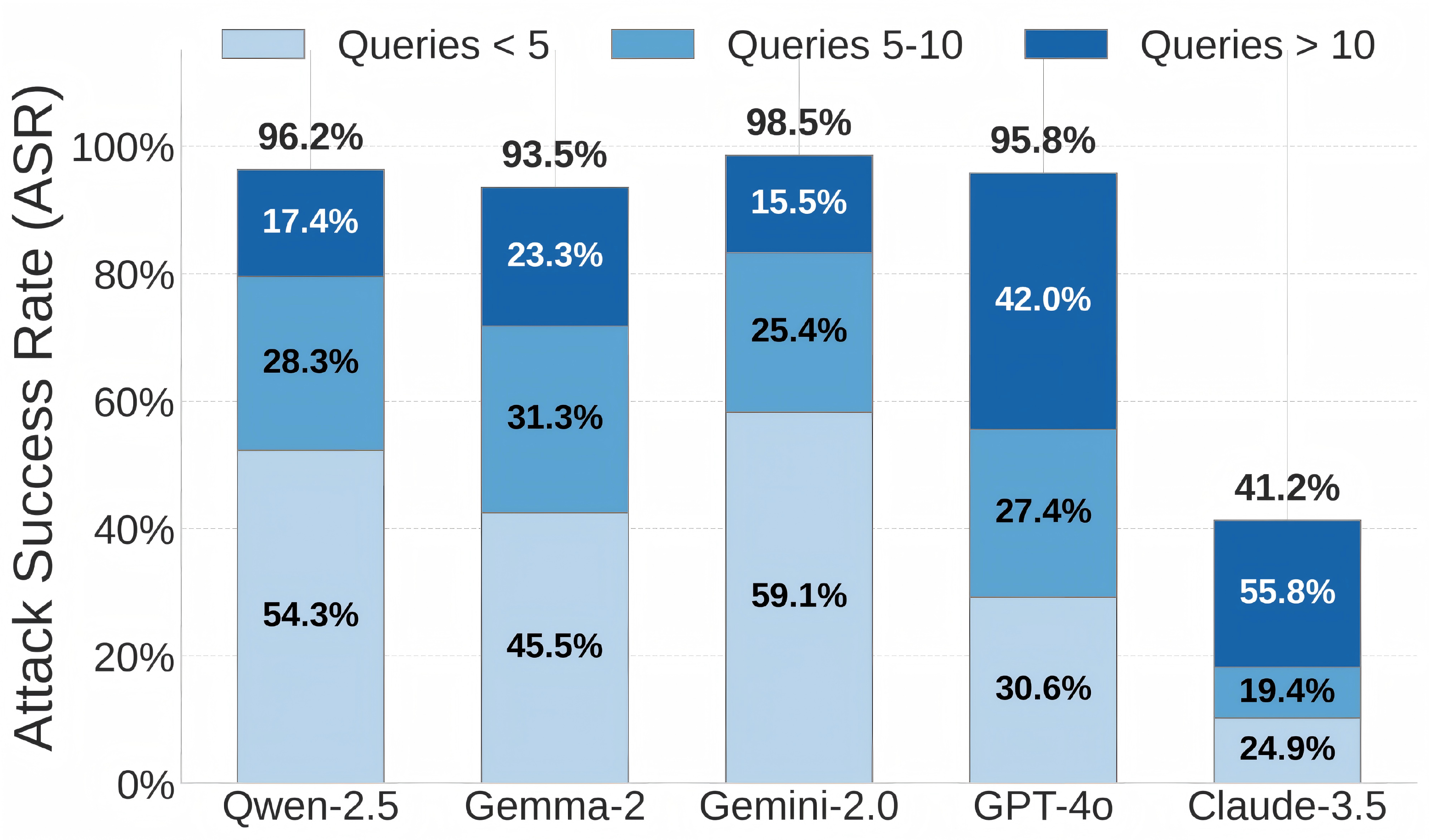}
    \vspace{-1.0em}
    \caption{Distribution of Query Counts in Successful Jailbreak Attempts.}
    \vspace{-1.0em}\label{fig:distribution2}
\end{figure}

% \vspace{-0.4em}
\subsection{Main Results}
% \vspace{-0.2em}

We conduct a comprehensive evaluation of MAJIC and SOTA baseline attack methods. The experimental results, detailed in Table~\ref{tab:jb-main} and Table~\ref{tab:jb-aqc-only}, unequivocally demonstrate MAJIC's superior performance in ASR, HS, and query efficiency across diverse LLMs. Detailed settings of our MAJIC can be found in the appendix.

\partitle{Superior Attack Effectiveness}
As shown in Table~\ref{tab:jb-main}, MAJIC consistently achieves the highest ASR and HS. On open-source models like Qwen-2.5-7b-it, MAJIC reaches 96.2\% ASR with a 0.55 HS, substantially outperforming the best baseline, Autodan-Turbo (with only 55.2\% ASR, 0.21 HS). A similar trend is also observed on Gemma-2-9b-it.
MAJIC's superiority is further amplified in challenging closed-source models. It achieves 98.5\% ASR on Gemini-2.0-flash and 95.7\% ASR on GPT-4o, again significantly surpassing all baselines in both ASR and HS. Most notably, MAJIC obtains a 41.2\% ASR and 0.21 HS on the highly resistant Claude-3.5-sonnet, where other methods  fail in most cases (\eg, ASRs typically <4\% and HS <0.02). 
% This highlights MAJIC's excellent robustness.

\partitle{Exceptional Query Efficiency}
Beyond its effectiveness, MAJIC demonstrates remarkable query efficiency (as Table~\ref{tab:jb-aqc-only}), a critical advantage for black-box attacks. MAJIC drastically reduces the Average Query Count (AQC) compared to all baselines. For instance, it requires only 7.5 queries on Qwen-2.5-7b-it and 6.3 on Gemini-2.0-flash, a $5\sim8$$\times$ reduction compared to the suboptimal baseline (\ie, Autodan-Turbo) with 42.5 and 32.2 queries, respectively. Even on GPT-4o and Claude-3.5-sonnet, MAJIC maintains a significant efficiency lead (13.1 and 29.5 queries, respectively) while still achieving high ASR. This substantial reduction in AQC underscores MAJIC's ability to rapidly converge to effective jailbreak prompts. The distribution of queries in successful jailbreak prompts in MAJIC, depicted in Figure~\ref{fig:distribution2}, also supports this claim. In most cases, MAJIC can succeed in attacking with fewer than 10 queries or even 5 queries.

We further conduct experiments to assess the generalizability of MAJIC-generated prompts across different models and a wide range of harm categories. Detailed results are provided in the appendix.

\begin{figure}[t]
    \centering
    \includegraphics[width=0.95\linewidth]{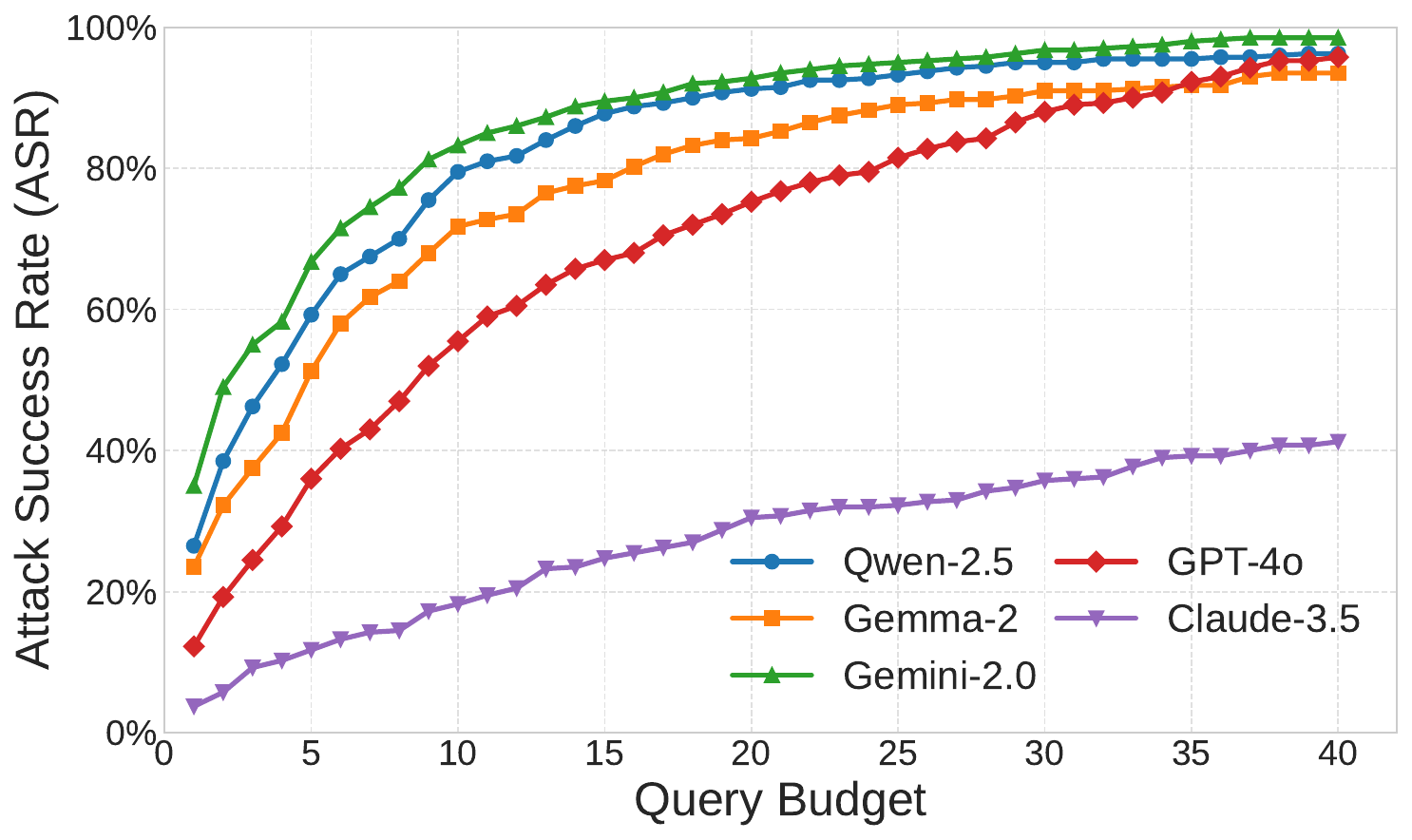}
    \vspace{-1.0em}
    \caption{MAJIC's ASR against different LLMs under various query budgets.}
    \vspace{-0.5em}\label{fig:query_budget_asr}
\end{figure}

% \begin{figure}[!t] % 外层的 figure 环境用于整体浮动，可以省略，但保留可以提供更灵活的浮动控制
%     \centering

%     \begin{minipage}[t]{0.48\linewidth}
%         \centering
%         \includegraphics[width=\linewidth]{figures/query_plot_2.pdf}
%         \vspace{-1.2em}
%         \caption{Distribution of queries in successful jailbreak prompts.}
%         \label{fig:distribution}
%     \end{minipage}%
%     \hfill
%     \begin{minipage}[t]{0.48\linewidth}
%         \centering
%         \includegraphics[width=\linewidth]{figures/budget_plot4.pdf}
%         \vspace{-1.2em}
%         \caption{MAJIC's ASR against different LLMs under various query budgets.} % 请替换为你真实的标题
%         \label{fig:query_budget_asr}
%     \end{minipage}

%     % 如果使用外层的 figure 环境，可以有一个总标题，但您希望每个图有独立标题，所以这里不需要
%     % \caption{Overall caption for the two figures side-by-side.}
%     % \label{fig:combined_figures_minipage}
%     \vspace{-1em}
% \end{figure}

% In summary, MAJIC establishes a new SOTA in black-box jailbreak attacks, delivering significantly higher ASR and HS with substantially fewer queries. Its robust performance, especially against highly secure models, showcases its advanced adaptive capabilities and practical threat potential.
% \vspace{-0.4em}
\subsection{Ablation Study}
\label{sec:ablation_study}
% \vspace{-0.2em}
We conducted ablation studies to assess the contributions of MAJIC's core components: the effectiveness of the Disguise Strategy Pool, Markov model-based strategy selection, matrix initialization and dynamic updates. Our results show that all proposed components are crucial for MAJIC's performance.

\begin{table}[t]
\centering
% \caption{Ablation on strategy selection. MAJIC's Markov model excels over fixed and random selections.}
% \label{tab:ablation-selection-singlecol}
% \vspace{-0.5em}
\scalebox{0.90}{
\begin{tabular}{l c c c}
\toprule
\multirow{2}{*}{\textbf{Method}} & \multirow{2}{*}{\textbf{ASR (\%)}} & \multirow{2}{*}{\textbf{HS}} & \multirow{2}{*}{\textbf{AQC}} \\
& & & \\
\midrule
GCG-T            & 21.5 & 0.09 & 64.7 \\
PAIR             & 32.7 & 0.10 & 52.9 \\
TAP              & 35.5 & 0.11 & 59.5 \\
PAP              & 34.7 & 0.14 & 55.6 \\
ReneLLM          & 44.0 & 0.17 & 50.2 \\
Autodan-Turbo    & 84.7 & 0.40 & 25.7 \\
\midrule
\rowcolor{yellow!30}
\textbf{R-MAJIC} & \textbf{65.2} & \textbf{0.25} & \textbf{33.4} \\
\rowcolor{yellow!30}
\textbf{F-MAJIC} & \textbf{68.5} & \textbf{0.26} & \textbf{31.6} \\
\rowcolor{yellow!30}
\textbf{MAJIC}   & \textbf{95.7} & \textbf{0.55} & \textbf{13.1} \\
\bottomrule
\end{tabular}}
\vspace{-0.5em}
\caption{Ablation on strategy selection. MAJIC's Markov model excels over fixed and random selections.}
\label{tab:ablation-selection-singlecol}
\vspace{-1.2em}
\end{table}

\begin{table}[t]
\centering
\scalebox{0.85}{
\begin{tabular}{l l c c c c}
\toprule
\multirow{2}{*}{\textbf{Variant}} & \textbf{Init.} & \textbf{Dyn.} & \textbf{ASR} & \multirow{2}{*}{\textbf{HS}} & \multirow{2}{*}{\textbf{AQC}} \\
                                 & \textbf{Matrix} & \textbf{Upd.} & \textbf{(\%)} & & \\
\midrule
MAJIC (--Init)   & Unif.           & $\checkmark$ & 70.3          & 0.32          & 26.5 \\
MAJIC (--DynUpd) & Lrn.            & $\times$     & 76.5          & 0.39          & 20.2 \\
\rowcolor{yellow!30}
\textbf{MAJIC}   & Lrn.            & $\checkmark$ & \textbf{95.7} & \textbf{0.55} & \textbf{13.1} \\
\bottomrule
\end{tabular}}
\caption{Ablation on matrix initialization and dynamic updates. ``Lrn.'' denotes a learned matrix; ``Unif.'' is uniform.}
\label{tab:ablation-init-dynamic-revised}
\vspace{-0.9em}
\end{table}

\partitle{Effectiveness of Disguise Strategy Pool}
% fixorder & random （only pool）vs baseline
To evaluate the effectiveness of the \textit{Disguise Strategy Pool}, we conduct ablation studies using simplified variants of MAJIC that omit the Markov-based strategy selection and dynamic updates. Specifically, we introduce two variants:\textbf{F-MAJIC} (strategies applied in a \underline{fixed} sequence) and \textbf{R-MAJIC} (strategies chosen \underline{randomly} upon failure).
Table~\ref{tab:ablation-selection-singlecol} shows results on GPT-4o (trends are consistent across other LLMs). Our Disguise Strategy Pool proves to be an important contributor. Even without adaptive selection mechanisms, both F-MAJIC and R-MAJIC achieve higher ASR and HS than most strong baselines, while also requiring less query cost.

\partitle{Impact of Markovian Strategy Selection Mechanism}
% To evaluate the efficacy of our Markov model for selecting disguise strategies, we compared MAJIC with two variants: \textbf{MAJIC-FixedOrder} (strategies applied in a fixed sequence) and \textbf{MAJIC-RandomOrder} (strategies chosen randomly upon failure). 
As shown in Table~\ref{tab:ablation-selection-singlecol}, MAJIC achieves an ASR of 95.7\%, significantly surpassesing F-MAJIC (68.5\%) and R-MAJIC (65.2\%). The HS and AQC also show substantial improvements with MAJIC's guided selection. This underscores the Markov model's vital role in constructing effective attack sequences by learning optimal strategy transitions, as opposed to simpler heuristic or random selection methods.

\partitle{Impact of Matrix Initialization and Dynamic Updates}
We then assess the importance of both the initial Markov transition matrix and its subsequent dynamic updates by comparing the full version of MAJIC against two variants. The first variant, denoted as \textbf{MAJIC (--DynUpd)}, utilizes the initialized Markov matrix but omits the dynamic updates during the attack phase. The second variant, named \textbf{MAJIC (--Init)}, performs dynamic updates but starts from a non-informative, uniform probability matrix instead of the one derived by our initialization mechanism. Results on GPT-4o(trends are consistent across other LLMs) are shown in Table~\ref{tab:ablation-init-dynamic-revised}. The full version of MAJIC, benefiting from both a good initial matrix ($\mathcal{M}_{L}$) and dynamic updates ($\checkmark$), achieving the highest performance (95.7\% ASR). Disabling dynamic updates (MAJIC (--DynUpd)) degrades ASR to 76.5\%, demonstrating the value of real-time adaptation. Similarly, disabling the initialization mechanism (MAJIC (--Init)), even with dynamic updates, results in a lower ASR of 70.3\% and a higher AQC (26.5). These results highlight that while dynamic updates can aid recovery from a suboptimal start, a well-constructed initial matrix provides a significant head start, leading to faster convergence and superior overall performance. Both components make contributions to MAJIC's success.

\partitle{Impact of Query Budgets}
We evaluate MAJIC's ASRs under varying query budgets (maximum query setting $N_{max}$) across five LLMs. The results, shown in Figure~\ref{fig:query_budget_asr}, reveal consistent trends: ASR improves as the query budget increases, but the rate of improvement diminishes beyond a certain threshold. For instance, MAJIC achieves a 95.75\% ASR on GPT-4o with a budget of 40 queries, compared to 12.25\% with only 1 query. Similar patterns are observed across other models, with Gemini-2.0 and Qwen-2.5 achieving 98.50\% and 96.20\% ASR, respectively, at their maximum budgets. Notably, Claude-3.5 exhibits the lowest ASR under all budgets, reflecting its more robust defenses, while GPT-4o shows a steeper improvement curve. These findings highlight MAJIC's ability to effectively adapt and succeed within a constrained query budget, achieving high ASR with relatively few interactions.

\section{Conclusion}
\label{sec:conclusion}

% 需要修改，需要加上POOL的重要性！

In this paper, we introduce MAJIC, a novel black-box jailbreak attack framework that leverages a dynamic Markov model to intelligently select and fuse attack strategies from an innovative disguise strategy pool. 
% Our core contribution lay in MAJIC's ability to adaptively learn optimal sequences of these strategies to bypass the safety mechanisms of LLMs. 
We conduct extensive experiments to evaluate MAJIC. Compared to existing baseline attacks, MAJIC achieves significantly higher Attack Success Rates and Harmfulness Scores with substantially fewer queries across a wide range of powerful closed-source and open-source LLMs. Ablation studies underscored the critical roles of our designed strategy pool, the proposed Markov model, the initialization mechanism for the Markov transition matrix, and the dynamic update algorithm in MAJIC. Furthermore, the evaluation results demonstrate MAJIC's strong attack generalizability across different models and its broad efficacy across various harm categories. 
The success of MAJIC underscores the ongoing challenges in robustly aligning LLMs and defending against complex jailbreak prompts. It highlights the need for the development of more advanced and holistic defense strategies that can anticipate and counter such dynamic, fusion-based attacks. 

\bibliography{main}

\begin{thebibliography}{47}
\providecommand{\natexlab}[1]{#1}

\bibitem[{Achiam et~al.(2023)Achiam, Adler, Agarwal, Ahmad, Akkaya, Aleman, Almeida, Altenschmidt, Altman, Anadkat et~al.}]{openai2024gpt4technicalreport}
Achiam, J.; Adler, S.; Agarwal, S.; Ahmad, L.; Akkaya, I.; Aleman, F.~L.; Almeida, D.; Altenschmidt, J.; Altman, S.; Anadkat, S.; et~al. 2023.
\newblock Gpt-4 technical report.
\newblock \emph{arXiv preprint arXiv:2303.08774}.

\bibitem[{Andriushchenko, Croce, and Flammarion(2024)}]{andriushchenko2024jailbreaking}
Andriushchenko, M.; Croce, F.; and Flammarion, N. 2024.
\newblock Jailbreaking leading safety-aligned llms with simple adaptive attacks.
\newblock \emph{arXiv preprint arXiv:2404.02151}.

\bibitem[{Andriushchenko and Flammarion(2024)}]{pasttense_attack}
Andriushchenko, M.; and Flammarion, N. 2024.
\newblock Does Refusal Training in LLMs Generalize to the Past Tense?
\newblock In \emph{NeurIPS Safe Generative AI Workshop}.

\bibitem[{{Anthropic}(2024)}]{claude3.5}
{Anthropic}. 2024.
\newblock Introducing Claude 3.5 Sonnet.
\newblock \url{https://www.anthropic.com/news/claude-3-5-sonnet}.
\newblock Accessed: 2024-6-21.

\bibitem[{Chao et~al.(2023)Chao, Robey, Dobriban, Hassani, Pappas, and Wong}]{pair}
Chao, P.; Robey, A.; Dobriban, E.; Hassani, H.; Pappas, G.~J.; and Wong, E. 2023.
\newblock Jailbreaking black box large language models in twenty queries.
\newblock In \emph{NeurIPS Workshop R0-FoMo}.

\bibitem[{Deng et~al.(2024)Deng, Zhang, Pan, and Bing}]{language_attack}
Deng, Y.; Zhang, W.; Pan, S.~J.; and Bing, L. 2024.
\newblock Multilingual Jailbreak Challenges in Large Language Models.
\newblock In \emph{International Conference on Learning Representations}.

\bibitem[{Ding et~al.(2024)Ding, Kuang, Ma, Cao, Xian, Chen, and Huang}]{renellm}
Ding, P.; Kuang, J.; Ma, D.; Cao, X.; Xian, Y.; Chen, J.; and Huang, S. 2024.
\newblock A Wolf in Sheep’s Clothing: Generalized Nested Jailbreak Prompts can Fool Large Language Models Easily.
\newblock In \emph{Conference of the North American Chapter of the Association for Computational Linguistics}, 2136--2153.

\bibitem[{Du et~al.(2025)Du, Mo, Wen, Gu, Zheng, Jin, and Shi}]{du2025multi}
Du, X.; Mo, F.; Wen, M.; Gu, T.; Zheng, H.; Jin, H.; and Shi, J. 2025.
\newblock Multi-turn jailbreaking large language models via attention shifting.
\newblock In \emph{Proceedings of the AAAI Conference on Artificial Intelligence}, volume~39, 23814--23822.

\bibitem[{Grattafiori et~al.(2024)Grattafiori, Dubey, Jauhri, Pandey, Kadian, Al-Dahle, Letman, Mathur, Schelten, Vaughan et~al.}]{llama3}
Grattafiori, A.; Dubey, A.; Jauhri, A.; Pandey, A.; Kadian, A.; Al-Dahle, A.; Letman, A.; Mathur, A.; Schelten, A.; Vaughan, A.; et~al. 2024.
\newblock The llama 3 herd of models.
\newblock \emph{arXiv preprint arXiv:2407.21783}.

\bibitem[{Guo et~al.(2025)Guo, Yang, Zhang, Song, Zhang, Xu, Zhu, Ma, Wang, Bi et~al.}]{deepseekr1}
Guo, D.; Yang, D.; Zhang, H.; Song, J.; Zhang, R.; Xu, R.; Zhu, Q.; Ma, S.; Wang, P.; Bi, X.; et~al. 2025.
\newblock Deepseek-r1: Incentivizing reasoning capability in llms via reinforcement learning.
\newblock \emph{arXiv preprint arXiv:2501.12948}.

\bibitem[{Guo et~al.(2024)Guo, Yu, Zhang, Qin, and Hu}]{coldattack}
Guo, X.; Yu, F.; Zhang, H.; Qin, L.; and Hu, B. 2024.
\newblock COLD-attack: jailbreaking LLMs with stealthiness and controllability.
\newblock In \emph{International Conference on Machine Learning}, 16974--17002.

\bibitem[{Hu, Chen, and Ho(2024)}]{NEURIPS2024jailbreak2}
Hu, X.; Chen, P.-Y.; and Ho, T.-Y. 2024.
\newblock Gradient Cuff: Detecting Jailbreak Attacks on Large Language Models by Exploring Refusal Loss Landscapes.
\newblock In \emph{Advances in Neural Information Processing Systems}, volume~37, 126265--126296.

\bibitem[{Hu, Chen, and Ho(2025)}]{hu2025token}
Hu, X.; Chen, P.-Y.; and Ho, T.-Y. 2025.
\newblock Token highlighter: Inspecting and mitigating jailbreak prompts for large language models.
\newblock In \emph{Proceedings of the AAAI Conference on Artificial Intelligence}, volume~39, 27330--27338.

\bibitem[{Huang, Li, and Tang(2024)}]{endless}
Huang, B.~R.; Li, M.; and Tang, L. 2024.
\newblock Endless Jailbreaks with Bijection Learning.
\newblock \emph{arXiv preprint arXiv:2410.01294}.

\bibitem[{Hurst et~al.(2024)Hurst, Lerer, Goucher, Perelman, Ramesh, Clark, Ostrow, Welihinda, Hayes, Radford et~al.}]{gpt4o}
Hurst, A.; Lerer, A.; Goucher, A.~P.; Perelman, A.; Ramesh, A.; Clark, A.; Ostrow, A.; Welihinda, A.; Hayes, A.; Radford, A.; et~al. 2024.
\newblock Gpt-4o system card.
\newblock \emph{arXiv preprint arXiv:2410.21276}.

\bibitem[{Jia et~al.(2024)Jia, Pang, Du, Huang, Gu, Liu, Cao, and Lin}]{imporved-gcg}
Jia, X.; Pang, T.; Du, C.; Huang, Y.; Gu, J.; Liu, Y.; Cao, X.; and Lin, M. 2024.
\newblock Improved techniques for optimization-based jailbreaking on large language models.
\newblock \emph{arXiv preprint arXiv:2405.21018}.

\bibitem[{Jiang et~al.(2023)Jiang, Sablayrolles, Mensch, Bamford, Chaplot, de~las Casas, Bressand, Lengyel, Lample, Saulnier, Lavaud, Lachaux, Stock, Scao, Lavril, Wang, Lacroix, and Sayed}]{mistral}
Jiang, A.~Q.; Sablayrolles, A.; Mensch, A.; Bamford, C.; Chaplot, D.~S.; de~las Casas, D.; Bressand, F.; Lengyel, G.; Lample, G.; Saulnier, L.; Lavaud, L.~R.; Lachaux, M.-A.; Stock, P.; Scao, T.~L.; Lavril, T.; Wang, T.; Lacroix, T.; and Sayed, W.~E. 2023.
\newblock Mistral 7B.
\newblock \emph{arXiv preprint arXiv:2310.06825}.

\bibitem[{Jin et~al.(2024)Jin, Zhou, Menke, and Wang}]{NEURIPS2024jailbreak1}
Jin, H.; Zhou, A.; Menke, J.~D.; and Wang, H. 2024.
\newblock Jailbreaking Large Language Models Against Moderation Guardrails via Cipher Characters.
\newblock In \emph{Advances in Neural Information Processing Systems}, volume~37, 59408--59435.

\bibitem[{Li et~al.(2025{\natexlab{a}})Li, Ye, Wu, Yan, Wang, and Li}]{jailpo}
Li, H.; Ye, J.; Wu, J.; Yan, T.; Wang, C.; and Li, Z. 2025{\natexlab{a}}.
\newblock JailPO: A Novel Black-box Jailbreak Framework via Preference Optimization against Aligned LLMs.
\newblock In \emph{Proceedings of the AAAI Conference on Artificial Intelligence}, volume~39, 27419--27427.

\bibitem[{Li et~al.(2024)Li, Zhou, Zhu, Yao, Liu, and Han}]{deepinception}
Li, X.; Zhou, Z.; Zhu, J.; Yao, J.; Liu, T.; and Han, B. 2024.
\newblock DeepInception: Hypnotize Large Language Model to Be Jailbreaker.
\newblock In \emph{NeurIPS Safe Generative AI Workshop}.

\bibitem[{Li et~al.(2025{\natexlab{b}})Li, Shao, He, Guo, Zhang, Qin, Chen, Backes, Torr, Tao et~al.}]{li2025rethinking}
Li, Y.; Shao, S.; He, Y.; Guo, J.; Zhang, T.; Qin, Z.; Chen, P.-Y.; Backes, M.; Torr, P.; Tao, D.; et~al. 2025{\natexlab{b}}.
\newblock Rethinking data protection in the (generative) artificial intelligence era.
\newblock \emph{arXiv preprint arXiv:2507.03034}.

\bibitem[{Lin et~al.(2025)Lin, Han, Li, and Liu}]{attacksunderstanding}
Lin, R.; Han, B.; Li, F.; and Liu, T. 2025.
\newblock Understanding and Enhancing the Transferability of Jailbreaking Attacks.
\newblock In \emph{International Conference on Learning Representations}.

\bibitem[{Liu et~al.(2025{\natexlab{a}})Liu, Li, Suh, Vorobeychik, Mao, Jha, McDaniel, Sun, Li, and Xiao}]{autodanturbo}
Liu, X.; Li, P.; Suh, E.; Vorobeychik, Y.; Mao, Z.; Jha, S.; McDaniel, P.; Sun, H.; Li, B.; and Xiao, C. 2025{\natexlab{a}}.
\newblock Autodan-turbo: A lifelong agent for strategy self-exploration to jailbreak llms.
\newblock In \emph{International Conference on Learning Representations}.

\bibitem[{Liu et~al.(2023)Liu, Deng, Xu, Li, Zheng, Zhang, Zhao, Zhang, Wang, and Liu}]{liu2023jailbreaking}
Liu, Y.; Deng, G.; Xu, Z.; Li, Y.; Zheng, Y.; Zhang, Y.; Zhao, L.; Zhang, T.; Wang, K.; and Liu, Y. 2023.
\newblock Jailbreaking chatgpt via prompt engineering: An empirical study.
\newblock \emph{arXiv preprint arXiv:2305.13860}.

\bibitem[{Liu et~al.(2025{\natexlab{b}})Liu, He, Xiong, Fu, Deng, and Hooi}]{liu2025flipattack}
Liu, Y.; He, X.; Xiong, M.; Fu, J.; Deng, S.; and Hooi, B. 2025{\natexlab{b}}.
\newblock FlipAttack: Jailbreak LLMs via Flipping.
\newblock In \emph{International Conference on Machine Learning}.

\bibitem[{Mazeika et~al.(2024)Mazeika, Phan, Yin, Zou, Wang, Mu, Sakhaee, Li, Basart, Li et~al.}]{harmbench}
Mazeika, M.; Phan, L.; Yin, X.; Zou, A.; Wang, Z.; Mu, N.; Sakhaee, E.; Li, N.; Basart, S.; Li, B.; et~al. 2024.
\newblock HarmBench: a standardized evaluation framework for automated red teaming and robust refusal.
\newblock In \emph{International Conference on Machine Learning}, 35181--35224.

\bibitem[{Mehrotra et~al.(2024)Mehrotra, Zampetakis, Kassianik, Nelson, Anderson, Singer, and Karbasi}]{tap}
Mehrotra, A.; Zampetakis, M.; Kassianik, P.; Nelson, B.; Anderson, H.; Singer, Y.; and Karbasi, A. 2024.
\newblock Tree of attacks: Jailbreaking black-box llms automatically.
\newblock \emph{Advances in Neural Information Processing Systems}, 37: 61065--61105.

\bibitem[{Nijkamp et~al.(2023)Nijkamp, Pang, Hayashi, Tu, Wang, Zhou, Savarese, and Xiong}]{nijkamp2023codegen}
Nijkamp, E.; Pang, B.; Hayashi, H.; Tu, L.; Wang, H.; Zhou, Y.; Savarese, S.; and Xiong, C. 2023.
\newblock CodeGen: An Open Large Language Model for Code with Multi-Turn Program Synthesis.
\newblock In \emph{International Conference on Learning Representations}.

\bibitem[{Norris(1998)}]{norris1998markov}
Norris, J.~R. 1998.
\newblock \emph{Markov chains}.
\newblock 2. Cambridge university press.

\bibitem[{Ramesh et~al.(2025)Ramesh, Bhardwaj, Saibewar, and Kaul}]{rameshefficient}
Ramesh, A.; Bhardwaj, S.; Saibewar, A.; and Kaul, M. 2025.
\newblock Efficient Jailbreak Attack sequences on Large Language Models via Multi-Armed Bandit-based Context switching.
\newblock In \emph{International Conference on Learning Representations}.

\bibitem[{Shen et~al.(2023)Shen, Jin, Huang, Liu, Dong, Guo, Wu, Liu, and Xiong}]{shen2023large}
Shen, T.; Jin, R.; Huang, Y.; Liu, C.; Dong, W.; Guo, Z.; Wu, X.; Liu, Y.; and Xiong, D. 2023.
\newblock Large language model alignment: A survey.
\newblock \emph{arXiv preprint arXiv:2309.15025}.

\bibitem[{Shen et~al.(2024)Shen, Chen, Backes, Shen, and Zhang}]{dan}
Shen, X.; Chen, Z.; Backes, M.; Shen, Y.; and Zhang, Y. 2024.
\newblock "do anything now": Characterizing and evaluating in-the-wild jailbreak prompts on large language models.
\newblock In \emph{ACM SIGSAC Conference on Computer and Communications Security}, 1671--1685.

\bibitem[{Souly et~al.(2024)Souly, Lu, Bowen, Trinh, Hsieh, Pandey, Abbeel, Svegliato, Emmons, Watkins et~al.}]{strongreject}
Souly, A.; Lu, Q.; Bowen, D.; Trinh, T.; Hsieh, E.; Pandey, S.; Abbeel, P.; Svegliato, J.; Emmons, S.; Watkins, O.; et~al. 2024.
\newblock A StrongREJECT for Empty Jailbreaks.
\newblock In \emph{ICLR 2024 Workshop on Reliable and Responsible Foundation Models}.

\bibitem[{Team et~al.(2023)Team, Anil, Borgeaud, Alayrac, Yu, Soricut, Schalkwyk, Dai, Hauth, Millican et~al.}]{gemini}
Team, G.; Anil, R.; Borgeaud, S.; Alayrac, J.-B.; Yu, J.; Soricut, R.; Schalkwyk, J.; Dai, A.~M.; Hauth, A.; Millican, K.; et~al. 2023.
\newblock Gemini: a family of highly capable multimodal models.
\newblock \emph{arXiv preprint arXiv:2312.11805}.

\bibitem[{Team et~al.(2024)Team, Riviere, Pathak, Sessa, Hardin, Bhupatiraju, Hussenot, Mesnard, Shahriari, Ram{\'e} et~al.}]{gemma2}
Team, G.; Riviere, M.; Pathak, S.; Sessa, P.~G.; Hardin, C.; Bhupatiraju, S.; Hussenot, L.; Mesnard, T.; Shahriari, B.; Ram{\'e}, A.; et~al. 2024.
\newblock Gemma 2: Improving open language models at a practical size.
\newblock \emph{arXiv preprint arXiv:2408.00118}.

\bibitem[{Wei, Liu, and Erichson(2024)}]{emoji-attack}
Wei, Z.; Liu, Y.; and Erichson, N.~B. 2024.
\newblock Emoji Attack: A Method for Misleading Judge LLMs in Safety Risk Detection.
\newblock \emph{arXiv preprint arXiv:2411.01077}.

\bibitem[{Xiao et~al.(2024)Xiao, Yang, Chen, and Chen}]{distractatttack}
Xiao, Z.; Yang, Y.; Chen, G.; and Chen, Y. 2024.
\newblock Distract large language models for automatic jailbreak attack.
\newblock \emph{arXiv preprint arXiv:2403.08424}.

\bibitem[{Xu, Liu, and Liu(2024)}]{xu2024bag}
Xu, Z.; Liu, F.; and Liu, H. 2024.
\newblock Bag of Tricks: Benchmarking of Jailbreak Attacks on LLMs.
\newblock In \emph{Conference on Neural Information Processing Systems Datasets and Benchmarks Track}.

\bibitem[{Yang et~al.(2024)Yang, Yang, Zhang, Hui, Zheng, Yu, Li, Liu, Huang, Wei et~al.}]{qwen2.5}
Yang, A.; Yang, B.; Zhang, B.; Hui, B.; Zheng, B.; Yu, B.; Li, C.; Liu, D.; Huang, F.; Wei, H.; et~al. 2024.
\newblock Qwen2. 5 technical report.
\newblock \emph{arXiv preprint arXiv:2412.15115}.

\bibitem[{Yao et~al.(2024)Yao, Duan, Xu, Cai, Sun, and Zhang}]{yao2024survey}
Yao, Y.; Duan, J.; Xu, K.; Cai, Y.; Sun, Z.; and Zhang, Y. 2024.
\newblock A survey on large language model (llm) security and privacy: The good, the bad, and the ugly.
\newblock \emph{High-Confidence Computing}, 100211.

\bibitem[{Yi et~al.(2024)Yi, Liu, Sun, Cong, He, Song, Xu, and Li}]{jailbreaksurvey}
Yi, S.; Liu, Y.; Sun, Z.; Cong, T.; He, X.; Song, J.; Xu, K.; and Li, Q. 2024.
\newblock Jailbreak attacks and defenses against large language models: A survey.
\newblock \emph{arXiv preprint arXiv:2407.04295}.

\bibitem[{Yuan et~al.(2024)Yuan, Jiao, Wang, Huang, He, Shi, and Tu}]{cipherchat}
Yuan, Y.; Jiao, W.; Wang, W.; Huang, J.-t.; He, P.; Shi, S.; and Tu, Z. 2024.
\newblock GPT-4 Is Too Smart To Be Safe: Stealthy Chat with LLMs via Cipher.
\newblock In \emph{International Conference on Learning Representations}.

\bibitem[{Zeng et~al.(2024)Zeng, Lin, Zhang, Yang, Jia, and Shi}]{pap}
Zeng, Y.; Lin, H.; Zhang, J.; Yang, D.; Jia, R.; and Shi, W. 2024.
\newblock How johnny can persuade llms to jailbreak them: Rethinking persuasion to challenge ai safety by humanizing llms.
\newblock In \emph{Annual Meeting of the Association for Computational Linguistics}, 14322--14350.

\bibitem[{Zhang, Zhang, and Foerster(2024)}]{ICML-defense-1}
Zhang, Z.; Zhang, Q.; and Foerster, J.~N. 2024.
\newblock {PARDEN}, Can You Repeat That? {D}efending against Jailbreaks via Repetition.
\newblock In \emph{International Conference on Machine Learning}, 60271--60287.

\bibitem[{Zhou, Li, and Wang(2024)}]{NEURIPS2024defense1}
Zhou, A.; Li, B.; and Wang, H. 2024.
\newblock Robust Prompt Optimization for Defending Language Models Against Jailbreaking Attacks.
\newblock In \emph{Advances in Neural Information Processing Systems}, volume~37, 40184--40211.

\bibitem[{Zhu et~al.(2024)Zhu, Zhang, An, Wu, Barrow, Wang, Huang, Nenkova, and Sun}]{autodan}
Zhu, S.; Zhang, R.; An, B.; Wu, G.; Barrow, J.; Wang, Z.; Huang, F.; Nenkova, A.; and Sun, T. 2024.
\newblock AutoDAN: interpretable gradient-based adversarial attacks on large language models.
\newblock In \emph{Conference on Language Modeling}.

\bibitem[{Zou et~al.(2023)Zou, Wang, Carlini, Nasr, Kolter, and Fredrikson}]{gcg}
Zou, A.; Wang, Z.; Carlini, N.; Nasr, M.; Kolter, J.~Z.; and Fredrikson, M. 2023.
\newblock Universal and transferable adversarial attacks on aligned language models.
\newblock \emph{arXiv preprint arXiv:2307.15043}.

\end{thebibliography}

%  ==============checklist================================
% \input{checklist} 
%  ==============checklist================================

\end{document}